# Which glass stability parameters can assess the glass-forming ability of oxide systems?

Jeanini Jiusti, Daniel R. Cassar, Edgar D. Zanotto

*Department of Materials Engineering, Federal University of São Carlos, Graduate Program in Materials Science and Engineering, São Carlos, SP, Brazil*

Abstract

Glass forming ability (GFA) is a property of utmost importance in glass science and technology. In this paper, we used a statistical methodology--involving bootstrap sampling and the Wilcoxon test--to find out which glass stability parameters can better predict the glass forming ability. We collected or measured the necessary data for twelve stoichiometric oxide glasses that underwent predominant heterogeneous nucleation (the most common case). We found that some GS parameters could predict the GFA of these oxide glasses quite well, whereas others perform poorly. Parameter $\mathbf{K_w}$ was the top ranked, closely followed by the $K_H$, $\gamma$, H', $\Delta T_{rg}$, and $K_{cr}$. Our results corroborate previous reports carried out using a smaller number of glasses, much less GS parameters, and less rigorous statistics. We also found that using $T_c$ instead of $T_x$ improved the predictive power of these parameters. Finally, the *Jezica*, the only parameter considered here that predicts the GFA without requiring the production of a glass piece (i.e., without relying on any crystallization information), ranked reasonably well in our analysis.

## 1 Introduction

Glass forming ability (GFA) – usually taken as the minimum cooling rate necessary to form a glass piece of a certain size - is a property of utmost importance in glass science and technology. Substances with low GFA require specific cooling methods and have limitations on the maximum size of the glass pieces that can be obtained. Therefore, applications of glasses are bound to their GFA, and a poor GFA can be a major obstacle for developing novel vitreous compositions.

From a purely scientific perspective, a non-crystalline substance below its liquidus temperature and with at least one stable crystalline nucleus is thermodynamically instable, as crystal growth in this condition takes place spontaneously [1]. Therefore, a rigorous definition of GFA must involve knowledge of the minimum cooling rate necessary to form a non-crystalline material without any stable crystalline nucleus. However, from a practical viewpoint, a non-crystalline substance is considered a glass if its crystalline fraction is below a certain threshold value, usually taken as $10^{-2}$ to $10^{-6}$. Therefore, the practical definition of GFA involves the knowledge of the minimum cooling rate, $R_c$, necessary to form a non-crystalline material having a crystalline fraction equal to the assumed threshold value. This is the definition we will use in this study.

One can estimate the $R_c$ (and thus the GFA) by building a TTT diagram using crystal growth data with the assumption that surface (heterogeneous) crystallization from a certain number of sites overshadows volumetric (homogeneous) crystallization. However, the experimental process of data acquisition at several temperatures necessary for this method is very time-consuming. Other methods, such as those reported in references [2–10], also have inherent difficulties, making the task of estimating $R_c$ laborious and subjected to large uncertainties [11].

In the past six decades, several glass <u>stability</u> parameters (GS, the resistance of a glass against crystallization upon heating) have been proposed to estimate the GFA [12–36]. These parameters are based on characteristic temperatures that can be measured by thermal analysis, such as glass transition ($T_g$), onset of crystallization ($T_x$), crystallization peak ($T_c$), and melting point ($T_m$). Measuring these temperatures is straightforward and much less time-consuming than constructing a TTT diagram. Because these characteristic temperatures are often determined during the heating path, it is more adequate to call all of them GS parameters instead of GFA parameters. However, a major problem with this procedure is that one has to make a glass sample to be able to measure the GS, hence it is not a truly predictive method.

Previous studies on the possible correlation between the GS parameters and the GFA of oxide glasses concluded that only some of them are related to the GFA [10,24,37–40]. These studies, however, included a limited number of GS parameters (up to 15) and glasses (up to 8), and focused mainly on the data correlation via the coefficient of determination, $R^2$.

In this paper, we expand the analysis to 35 parameters proposed in the literature, perform the calculations using a larger number of glasses (12) and apply a more rigorous statistical procedure to test which parameters best *predict* the $R_c$ (which is different from which parameter best *correlates* with the $R_c$). This will be explained in Section 3.2. It is important to emphasize that the word "predict" is used here in the statistical sense. All GS parameters depend on information acquired from calorimetry experiments using glass samples, which require the production of a glass piece.

Although not a "glass stability" parameter itself, we also included in our analysis a recently proposed parameter denominated Jezica [11], $\eta(T_l)/T_l^2$, where $\eta(T_l)$ is the shear viscosity at the *liquidus, $T_l$*. This parameter depends only on properties of the equilibrium liquid instead of the characteristic temperatures determined from a previously prepared glass. Thus, *Jezica* is a parameter that can predict $R_c$ without the need to make the glass, which sets it apart from the other parameters considered in this work. For the sake of simplicity, from now on we refer to "the GS parameters" instead of "the GS parameters and Jezica".

## 2 Equations for GS and GFA

The GS parameters tested here and their respective references are shown in Table 1 with the notations regarding the melting temperature and the *liquidus* temperature ($T_m$ and $T_l$) from the original source. In this paper, however, we are considering only the so-called stoichiometric glasses, that is, glasses that are isochemical with their stable crystalline phase.

Table 1 - GS parameters used in this study. Some parameters appear twice in the table, one using $T_c$ and the other using $T_x$. We indicate which temperature is used between parentheses after the parameter name.

| Parameter | Ref | Parameter | Ref | Parameter | Ref |
|---|---|---|---|---|---|
| $T_{rg} = \dfrac{T_g}{T_m}$ | [12] | $K_H(T_x) = \dfrac{T_x - T_g}{T_m - T_x}$ | [13] | $K_H(T_c) = \dfrac{T_c - T_g}{T_m - T_c}$ | [24]* |
| $\Delta T_x = T_x - T_g$ | [30] | $\Delta T_c = T_c - T_g$ | [24]* | $K_S(T_c) = \dfrac{(T_c - T_x)(T_c - T_g)}{T_g}$ | [31] |
| $K_S(T_x) = \dfrac{(T_c - T_x)(T_x - T_g)}{T_g}$ | [24]* | $H'(T_x) = \dfrac{T_x - T_g}{T_g}$ | [31]** | $H'(T_c) = \dfrac{T_c - T_g}{T_g}$ | * |
| $\alpha(T_c) = \dfrac{T_c}{T_l}$ | [32] | $\alpha(T_x) = \dfrac{T_x}{T_l}$ | [24]* | $K_w(T_c) = \dfrac{T_c - T_g}{T_m}$ | [33] |
| $K_w(T_x) = \dfrac{T_x - T_g}{T_m}$ | [24]* | $K_{w2} = \dfrac{(T_c - T_x)(T_x - T_g)}{T_m}$ | [33] | $\Delta T_g = T_l - T_g$ | [24] |
| $\gamma(T_x) = \dfrac{T_x}{T_g + T_l}$ | [34] | $\gamma(T_c) = \dfrac{T_c}{T_g + T_l}$ | [34] | $\Delta T_l = T_l - T_x$ | [35] |
| $\beta = \dfrac{T_x}{T_g} + \dfrac{T_g}{T_l}$ | [36] | $\delta = \dfrac{T_x}{T_l - T_g}$ | [14] | $\phi = T_{rg}\left(\dfrac{\Delta T_x}{T_g}\right)^{0.143}$ | [15] |
| $\gamma_m = \dfrac{2T_x - T_g}{T_l}$ | [16] | $\xi = \dfrac{T_g}{T_l} + \dfrac{\Delta T_x}{T_x}$ | [17] | $\Delta T_{rg} = \dfrac{T_x - T_g}{T_l - T_g}$ | [17]** |
| $\beta_1 = \dfrac{T_x T_g}{(T_l - T_x)^2}$ | [18] | $\omega_0 = \dfrac{T_g}{T_x} - \dfrac{2T_g}{T_g + T_l}$ | [19] | $\omega_1 = \dfrac{T_g/(T_x - 2T_g)}{T_g + T_l}$ | [20] |
| $\omega_2 = \dfrac{T_g}{2T_x - T_g} - \dfrac{T_g}{T_l}$ | [21] | $\omega_3 = \dfrac{T_l(T_l + T_x)}{T_x(T_l - T_x)}$ | [22] | $\theta = \dfrac{T_x + T_g}{T_l\left(\dfrac{T_x - T_g}{T_l}\right)^{0{,}0728}}$ | [23] |
| $\gamma_c = \dfrac{3T_x - 2T_g}{T_l}$ | [25] | $\beta_2 = \dfrac{T_g}{T_x} - \dfrac{T_g}{1.3 T_l}$ | [26] | $K_{CR} = \dfrac{T_l - T_x}{T_l - T_g}$ | [27] |
| $G_p = \dfrac{(T_x - T_g) T_g}{(T_l - T_x)^2}$ | [28] | $K_m = \dfrac{(T_x - T_g)^2}{T_g}$ | [29] | | |

*Variation of other original equation, ** Cited in this reference without informing the source.

Most of the parameters considered in this work were deduced by the original authors as a proxy for GFA (except $K_S$, $\Delta T_x$, H', $\Delta T_g$, $K_w$, $K_{w2}$, and $K_m$), but all of them require information collected by carrying out thermal experiments on heating a glass piece or powder. Therefore, a glass must be made for their determination. Hence, as we already mentioned, we consider all the equations shown in Table 1 as **glass stability** parameters, and not GFA parameters. The only exception is the *Jezica*, which is a phenomenological equation that can predict GFA without assessing any property that would require a glass to be made in the first place. In other words, Jezica is a predictor that only requires two properties of the liquid, the viscosity and the liquidus. To compute the *Jezica,* we used the VFT expression, (Eq. 1) [41–43], to fit the collected viscosity data from several authors.

$$log_{10}(\eta) = A + \frac{B}{T-T_0}, \tag{1}$$

where A is the pre-exponential factor, B is an effective activation energy and $T_0$ is the temperature of the divergence of the model. All three parameters are adjusted to the experimental data.

Critical cooling rates, $R_c$, can be computed from the nose shaped TTT curves, which has been extensively studied [5,6,44]. For surface (heterogeneous) nucleation, the nose temperature ($T_n$) is that of maximum growth rate, $T(U_{max})$. Therefore, the GFA can be computed with Eq. 2. $T_{Umax}$ is the temperature of maximum growth rate and $t_n$ is the time at the nose. $t_n$ can be calculated from Eq. 3, where $X_s$ is the maximum allowed crystallized fraction, $N_s$ is the number of sites per unit area inducing surface crystallization, and $U_{max}$ is the maximum growth rate.

$$GFA = \frac{1}{log(R_c)} = \frac{1}{log\left(\frac{T_m - T_{U_{max}}}{t_n}\right)} \tag{2}$$

$$t_n = \sqrt{\frac{X_s}{\pi N_s U_{max}^2}} \tag{3}$$

In this study, we use the logarithm scale to cope with the wide variation of $R_c$ among the 12 substances studied.

## 3 Materials and methods

### 3.1 Experimental data

Table 2 shows all the relevant data used in this study. The crystal growth data necessary to compute the critical cooling rate, and the viscosity coefficients required to compute the *Jezica* parameter were collected from several references, cited in Table 2. The characteristic temperatures $T_g$, $T_x$, and $T_c$ were collected from Refs. [24,45,46], and experimentally determined in this paper for $BaO \cdot 2B_2O_3$ and $SrO \cdot 2B_2O_3$ using a Netzsch DSC 404 cell with a DTA sensor (the data is shown in S1). All data collected from the literature and measured in this study were determined in similar conditions for fine powders (<60 μm diameter), on heating at 10 K/min, and in atmospheric air. The $T_l$ were collected from the respective phase equilibrium diagrams. The dataset consisted of six silicate, five borate and germania ($GeO_2$) glasses.

Table 2 - Characteristic temperatures ($T_g$, $T_x$, $T_c$ and $T_l$), $U_{max}$, $T(U_{max})$ and parameters of VFT viscosity equation $\log_{10}(\eta) = A + B/(T-T_0)$ used in this study.

| Glass | $T_g$ (K) | $T_x$ (K) | $T_c$ (K) | $T_l$ (K) | $U_{max}$ (m.s$^{-1}$) | $T(U_{max})$ (K) | A | B | $T_0$ | Ref. |
|---|---|---|---|---|---|---|---|---|---|---|
| Li$_2$O·2B$_2$O$_3$ | 764 | 821 | 830 | 1190 | $2.9 \times 10^{-3}$ | 1114 | −3.74 | 1685 | 657.3 | [24,47–49] |
| Na$_2$O·2B$_2$O$_3$ | 730 | 813 | 825 | 1016 | $3.34 \times 10^{-5}$ | 972 | −3.03 | 1279 | 677.1 | [45,48,50,51] |
| SrO·2B$_2$O$_3$ | 900 | 1004 | 1023 | 1264 | $1.60 \times 10^{-4}$ | 1215 | −3.48 | 1614 | 817.7 | [this study] [48,52] |
| BaO·2B$_2$O$_3$ | 866 | 972 | 988 | 1181 | $4.31 \times 10^{-5}$ | 1083 | −5.01 | 2816 | 686.0 | [this study], [48,53–55] |
| PbO·2B$_2$O$_3$ | 710 | 826 | 866 | 1047 | $2.10 \times 10^{-6}$ | 979 | −3.74 | 1685 | 657.3 | [46,48,56] |
| GeO$_2$ | 819 | 1090 | 1174 | 1388 | $9.33 \times 10^{-8}$ | 1327 | −7.21 | 17516 | 48.9 | [24,57–60] |
| PbO·SiO$_2$ | 677 | 859 | 901 | 1037 | $5.11 \times 10^{-7}$ | 936 | −2.69 | 1899 | 556.0 | [24,61–65] |
| Li$_2$O·2SiO$_2$ | 740 | 819 | 886 | 1303 | $6.87 \times 10^{-5}$ | 1210 | −2.40 | 3082 | 509.6 | [24,66–74] |
| Na$_2$O·2SiO$_2$ | 713 | 897 | 918 | 1148 | $9.92 \times 10^{-7}$ | 1087 | −3.00 | 4254 | 429.3 | [24,63,75,76] |
| 2MgO·2Al$_2$O$_3$·5SiO$_2$ | 1072 | 1203 | 1238 | 1740 | $9.05 \times 10^{-6}$ | 1533 | −3.97 | 5316 | 762.0 | [24,77–80] |
| CaO·Al$_2$O$_3$·2SiO$_2$ | 1127 | 1280 | 1301 | 1833 | $1.48 \times 10^{-4}$ | 1664 | −3.32 | 3939 | 866.9 | [24,81–86] |
| CaO·MgO·2SiO$_2$ | 988 | 1148 | 1187 | 1664 | $2.21 \times 10^{-4}$ | 1614 | −4.79 | 4875 | 689.9 | [24,81,83,87–89] |

## 3.2 Statistical methods

Here we briefly discuss *bootstrapping*, the widespread statistical method [90–92] that we used to answer the main question of this work: "*which GS parameters best predict the GFA?*". For those unfamiliar with this procedure, a step-by-step description can be found in Section S2 of the Supplementary Material.

First, we generated 10,000 bootstrap samples (or replicates) from the dataset shown in Table 2. Due to the random nature of the drawing process with reposition of this step, it is expected that these bootstrap samples will not be equal to the original dataset, thus containing more than one instance of certain glasses and not containing any instances of other glasses. Here we call "unselected glasses" the set of glasses that are not present in a bootstrap sample.

Random noise was added to the bootstrap samples and to all the sets of "unselected glasses" aiming to simulate experimental uncertainty. This process is discussed in detail in the Supplementary Material.

For each bootstrap sample and each respective set of "unselected glasses", we computed

i.  $R_c$ using Eq. 2, with $X_s = 10^{-2}$ and $N_s = 10^3$ (we also considered a noise in the value of $\log_{10}(N_s)$, see the Supplementary Material);

ii. viscosity using Eq. 1 (only needed for the Jezica parameter); and

iii. all GS parameters using the equations shown in Table 1.

For each GS parameter, we performed a linear regression with the independent variable in which the GS parameter and the dependent variable are the base-10 logarithm of $R_c$. For each bootstrap sample generated, we had one linear regression for each GS parameter that was tested. We then used each of these linear regressions to predict the $\log_{10}(R_c)$ of the glasses that were left out of their respective bootstrap sample, the "unselected glasses". Having the calculated $\log_{10}(R_c)$ and the predicted $\log_{10}(R_c)$ of these unselected samples, we computed the absolute residual of the prediction, which is the absolute of the difference between them. All the absolute residuals of prediction were stored in a sequential list, one list for each GS parameter.

To test which GS parameter yields the lowest values of absolute residuals, we used the non-parametric Wilcoxon test with a statistical confidence of 99%. This test must be done in pairs of GS parameters. First, we tested the null hypothesis if the difference between the absolute residuals between a pair of GS parameters followed a symmetric distribution around zero. If we cannot reject the null hypothesis, then we cannot support that there is a statistical difference in the prediction of $\log_{10}(R_c)$ between the considered parameters. However, when the null hypothesis was rejected, we ran the Wilcoxon test again, this time testing a null hypothesis that the absolute residuals of one of the GS parameters is lower than the other. Using the Wilcoxon test with each possible pair of parameters, we can check which parameter is best for predicting the $\log_{10}(R_c)$, which will have the lowest absolute residual values.

## 4   Results and discussion

### 4.1 Statistical methods

Table 3 shows the results of the Wilcoxon tests for all possible pairs of GS parameters. Number "−1" indicates that the parameter in the column is better than the one in the line. Number "1" indicates that the parameter in the line is better than the one in the column, whereas "0" indicates that the parameters do not differ in terms of predicting the $\log_{10}(R_c)$. The reader can observe a diagonal of "zeros" (highlighted in the table) that refers to the comparison between the parameter with itself. In Table 3, the parameters are ranked from the best to the worst (left to right).

Overall, with the proposed numerical and statistical procedure, the best GS expression was the **$K_w$ ($T_c$) = ($T_c − T_g$)/$T_l$**. The runner up GS parameters were $K_w(T_x)$, $\gamma(T_c)$, H'($T_x$), $K_H(T_c)$, H'($T_c$), $K_H(T_x)$, $\Delta T_{rg}$, and $K_{cr}$. Nascimento et al. [24] also concluded that $K_w$, $K_H$, and $\gamma$ indeed show a good correlation with the GFA of oxide glasses. In this paper, even including many other GFA expressions and testing them for statistical prediction ability, $K_w$, $K_H$, and $\gamma$ still made the podium, sharing their top positions with H' (which was not tested in [24]). The $K_H$ parameter was revised by Kozmidis-Petrovic and Šestác [40] who pointed out that this is indeed one of the best to estimate the glass stability and vitrification ability.

The *Jezica* parameter, in the eleventh position in the ranking, is the only parameter that can be used to phenomenologically (in addition to statistically) predict GFA without obtaining a glass. It only requires knowledge of the liquid viscosity at the *liquidus* and the respective *liquidus* temperature. This is a relevant result because no direct information on crystallization kinetics is required to use this parameter.

Apart from the *Jezica*, the only other parameter that does not consider the crystallization peak or the crystallization onset is the $T_{gr}$. However, it was the worst performer in our analysis. $T_{gr}$ was the first glass stability parameter proposed, and it was deduced considering only internal homogeneous nucleation. However, the GFA computed in this work (see Eq. 2) was obtained with the assumption that surface heterogeneous crystal nucleation prevails over homogeneous nucleation. Due to this, it is no surprise that $T_{gr}$ did not perform well in our analysis. For metallic glasses, where copious homogeneous nucleation is predominant, this parameter usually shows reasonable results, although it is often not listed among the best [18,19,93].

Due to the random nature of the bootstrap samples, a visualization of the employed statistical method is not straightforward. Figure 1a and 1c were plotted aiming to provide a better understanding of this procedure. We will call these figures "bootstrap distribution plots", and they were made for the GS parameters that scored the best (Figure 1a) and the worst (Figure 1c) in our analysis, $K_w(T_c)$ and $T_{gr}$, respectively.

Table 3 - Wilcoxon test results. "0" indicates that the parameters are not different, "1" indicates the parameter in the line is preferable and "-1" indicates the parameters in the column is preferable. The R² mode (the most frequent R² of linear regression of the bootstrap samples) for each parameter is also shown

| | $K_w(T_c)$ | $K_w(T_x)$ | $\gamma(T_c)$ | $H'(T_x)$ | $K_H(T_c)$ | $H'(T_c)$ | $K_H(T_x)$ | $\Delta T_{gr}$ | $K_{cr}$ | $\gamma_c$ | JZCA | $K_m$ | $\omega_2$ | $\gamma_m$ | $\beta$ | $\omega_0$ | $\omega_1$ | $\gamma(T_x)$ | $\beta_2$ | $G_p$ | $\Delta T_c$ | $\Delta T_x$ | $K_{w2}$ | $\alpha(T_c)$ | $K_S(T_x)$ | $\xi$ | $K_S(T_c)$ | $\alpha(T_x)$ | $\Delta T_l$ | $\omega_3$ | $\theta$ | $\phi$ | $\beta_1$ | $\Delta T_g$ | $\delta$ | $T_{gr}$ | R²mode |
|---|---|---|---|---|---|---|---|---|---|---|---|---|---|---|---|---|---|---|---|---|---|---|---|---|---|---|---|---|---|---|---|---|---|---|---|---|---|
| $K_w(T_c)$ | 0 | 1 | 1 | 1 | 1 | 1 | 1 | 1 | 1 | 1 | 1 | 1 | 1 | 1 | 1 | 1 | 1 | 1 | 1 | 1 | 1 | 1 | 1 | 1 | 1 | 1 | 1 | 1 | 1 | 1 | 1 | 1 | 1 | 1 | 1 | 1 | **0.79** |
| $K_w(T_x)$ | -1 | 0 | 1 | 1 | 1 | 1 | 1 | 1 | 1 | 1 | 1 | 1 | 1 | 1 | 1 | 1 | 1 | 1 | 1 | 1 | 1 | 1 | 1 | 1 | 1 | 1 | 1 | 1 | 1 | 1 | 1 | 1 | 1 | 1 | 1 | 1 | **0.77** |
| $\gamma(T_c)$ | -1 | -1 | 0 | 0 | 1 | 1 | 1 | 1 | 1 | 1 | 1 | 1 | 1 | 1 | 1 | 1 | 1 | 1 | 1 | 1 | 1 | 1 | 1 | 1 | 1 | 1 | 1 | 1 | 1 | 1 | 1 | 1 | 1 | 1 | 1 | 1 | **0.70** |
| $H'(T_x)$ | -1 | -1 | 0 | 0 | 0 | 1 | 1 | 1 | 1 | 1 | 1 | 1 | 1 | 1 | 1 | 1 | 1 | 1 | 1 | 1 | 1 | 1 | 1 | 1 | 1 | 1 | 1 | 1 | 1 | 1 | 1 | 1 | 1 | 1 | 1 | 1 | **0.78** |
| $K_H(T_c)$ | -1 | -1 | -1 | 0 | 0 | 0 | 1 | 1 | 1 | 1 | 1 | 1 | 1 | 1 | 1 | 1 | 1 | 1 | 1 | 1 | 1 | 1 | 1 | 1 | 1 | 1 | 1 | 1 | 1 | 1 | 1 | 1 | 1 | 1 | 1 | 1 | **0.71** |
| $H'(T_c)$ | -1 | -1 | -1 | -1 | 0 | 0 | 1 | 1 | 1 | 1 | 1 | 1 | 1 | 1 | 1 | 1 | 1 | 1 | 1 | 1 | 1 | 1 | 1 | 1 | 1 | 1 | 1 | 1 | 1 | 1 | 1 | 1 | 1 | 1 | 1 | 1 | **0.77** |
| $K_H(T_x)$ | -1 | -1 | -1 | -1 | -1 | -1 | 0 | 1 | 1 | 1 | 1 | 1 | 1 | 1 | 1 | 1 | 1 | 1 | 1 | 1 | 1 | 1 | 1 | 1 | 1 | 1 | 1 | 1 | 1 | 1 | 1 | 1 | 1 | 1 | 1 | 1 | **0.72** |
| $\Delta T_{gr}$ | -1 | -1 | -1 | -1 | -1 | -1 | -1 | 0 | 0 | 1 | 1 | 1 | 1 | 1 | 1 | 1 | 1 | 1 | 1 | 1 | 1 | 1 | 1 | 1 | 1 | 1 | 1 | 1 | 1 | 1 | 1 | 1 | 1 | 1 | 1 | 1 | **0.72** |
| $K_{cr}$ | -1 | -1 | -1 | -1 | -1 | -1 | -1 | 0 | 0 | 1 | 1 | 1 | 1 | 1 | 1 | 1 | 1 | 1 | 1 | 1 | 1 | 1 | 1 | 1 | 1 | 1 | 1 | 1 | 1 | 1 | 1 | 1 | 1 | 1 | 1 | 1 | **0.72** |
| $\gamma_c$ | -1 | -1 | -1 | -1 | -1 | -1 | -1 | -1 | -1 | 0 | 1 | 1 | 1 | 1 | 1 | 1 | 1 | 1 | 1 | 1 | 1 | 1 | 1 | 1 | 1 | 1 | 1 | 1 | 1 | 1 | 1 | 1 | 1 | 1 | 1 | 1 | 0.65 |
| JZCA | -1 | -1 | -1 | -1 | -1 | -1 | -1 | -1 | -1 | -1 | 0 | 1 | 1 | 1 | 1 | 1 | 1 | 1 | 1 | 1 | 1 | 1 | 1 | 1 | 1 | 1 | 1 | 1 | 1 | 1 | 1 | 1 | 1 | 1 | 1 | 1 | 0.63 |
| $K_m$ | -1 | -1 | -1 | -1 | -1 | -1 | -1 | -1 | -1 | -1 | -1 | 0 | 0 | 1 | 1 | 1 | 1 | 1 | 1 | 1 | 1 | 1 | 1 | 1 | 1 | 1 | 1 | 1 | 1 | 1 | 1 | 1 | 1 | 1 | 1 | 1 | 0.64 |
| $\omega_2$ | -1 | -1 | -1 | -1 | -1 | -1 | -1 | -1 | -1 | -1 | -1 | 0 | 0 | 1 | 1 | 1 | 1 | 1 | 1 | 1 | 1 | 1 | 1 | 1 | 1 | 1 | 1 | 1 | 1 | 1 | 1 | 1 | 1 | 1 | 1 | 1 | 0.62 |
| $\gamma_m$ | -1 | -1 | -1 | -1 | -1 | -1 | -1 | -1 | -1 | -1 | -1 | -1 | -1 | 0 | 1 | 1 | 1 | 1 | 1 | 1 | 1 | 1 | 1 | 1 | 1 | 1 | 1 | 1 | 1 | 1 | 1 | 1 | 1 | 1 | 1 | 1 | 0.61 |
| $\beta$ | -1 | -1 | -1 | -1 | -1 | -1 | -1 | -1 | -1 | -1 | -1 | -1 | -1 | -1 | 0 | 1 | 1 | 1 | 1 | 1 | 1 | 1 | 1 | 1 | 1 | 1 | 1 | 1 | 1 | 1 | 1 | 1 | 1 | 1 | 1 | 1 | 0.55 |
| $\omega_0$ | -1 | -1 | -1 | -1 | -1 | -1 | -1 | -1 | -1 | -1 | -1 | -1 | -1 | -1 | -1 | 0 | 0 | 1 | 1 | 1 | 1 | 1 | 1 | 1 | 1 | 1 | 1 | 1 | 1 | 1 | 1 | 1 | 1 | 1 | 1 | 1 | 0.59 |
| $\omega_1$ | -1 | -1 | -1 | -1 | -1 | -1 | -1 | -1 | -1 | -1 | -1 | -1 | -1 | -1 | -1 | 0 | 0 | 0 | 1 | 1 | 1 | 1 | 1 | 1 | 1 | 1 | 1 | 1 | 1 | 1 | 1 | 1 | 1 | 1 | 1 | 1 | 0.50 |
| $\gamma(T_x)$ | -1 | -1 | -1 | -1 | -1 | -1 | -1 | -1 | -1 | -1 | -1 | -1 | -1 | -1 | -1 | -1 | 0 | 0 | 1 | 1 | 1 | 1 | 1 | 1 | 1 | 1 | 1 | 1 | 1 | 1 | 1 | 1 | 1 | 1 | 1 | 1 | 0.53 |
| $\beta_2$ | -1 | -1 | -1 | -1 | -1 | -1 | -1 | -1 | -1 | -1 | -1 | -1 | -1 | -1 | -1 | -1 | -1 | -1 | 0 | 0 | 1 | 1 | 1 | 1 | 1 | 1 | 1 | 1 | 1 | 1 | 1 | 1 | 1 | 1 | 1 | 1 | 0.59 |
| $G_p$ | -1 | -1 | -1 | -1 | -1 | -1 | -1 | -1 | -1 | -1 | -1 | -1 | -1 | -1 | -1 | -1 | -1 | -1 | 0 | 0 | 0 | 1 | 1 | 1 | 1 | 1 | 1 | 1 | 1 | 1 | 1 | 1 | 1 | 1 | 1 | 1 | 0.53 |
| $\Delta T_c$ | -1 | -1 | -1 | -1 | -1 | -1 | -1 | -1 | -1 | -1 | -1 | -1 | -1 | -1 | -1 | -1 | -1 | -1 | -1 | 0 | 0 | 1 | 0 | 1 | 1 | 1 | 1 | 1 | 1 | 1 | 1 | 1 | 1 | 1 | 1 | 1 | 0.62 |
| $\Delta T_x$ | -1 | -1 | -1 | -1 | -1 | -1 | -1 | -1 | -1 | -1 | -1 | -1 | -1 | -1 | -1 | -1 | -1 | -1 | -1 | -1 | -1 | 0 | 0 | 1 | 1 | 1 | 1 | 1 | 1 | 1 | 1 | 1 | 1 | 1 | 1 | 1 | 0.61 |
| $K_{w2}$ | -1 | -1 | -1 | -1 | -1 | -1 | -1 | -1 | -1 | -1 | -1 | -1 | -1 | -1 | -1 | -1 | -1 | -1 | -1 | -1 | 0 | 0 | 0 | -1 | 1 | 1 | 1 | 1 | 1 | 1 | 1 | 1 | 1 | 1 | 1 | 1 | 0.55 |
| $\alpha(T_c)$ | -1 | -1 | -1 | -1 | -1 | -1 | -1 | -1 | -1 | -1 | -1 | -1 | -1 | -1 | -1 | -1 | -1 | -1 | -1 | -1 | -1 | -1 | 1 | 0 | 1 | 1 | 1 | 1 | 1 | 1 | 1 | 1 | 1 | 1 | 1 | 1 | 0.48 |
| $K_S(T_x)$ | -1 | -1 | -1 | -1 | -1 | -1 | -1 | -1 | -1 | -1 | -1 | -1 | -1 | -1 | -1 | -1 | -1 | -1 | -1 | -1 | -1 | -1 | -1 | -1 | 0 | 0 | 1 | 1 | 1 | 1 | 1 | 1 | 1 | 1 | 1 | 1 | 0.53 |
| $\xi$ | -1 | -1 | -1 | -1 | -1 | -1 | -1 | -1 | -1 | -1 | -1 | -1 | -1 | -1 | -1 | -1 | -1 | -1 | -1 | -1 | -1 | -1 | -1 | -1 | 0 | 0 | 1 | 1 | 1 | 1 | 1 | 1 | 1 | 1 | 1 | 1 | 0.44 |
| $K_S(T_c)$ | -1 | -1 | -1 | -1 | -1 | -1 | -1 | -1 | -1 | -1 | -1 | -1 | -1 | -1 | -1 | -1 | -1 | -1 | -1 | -1 | -1 | -1 | -1 | -1 | -1 | -1 | 0 | 1 | 1 | 1 | 1 | 1 | 1 | 1 | 1 | 1 | 0.00 |
| $\alpha(T_x)$ | -1 | -1 | -1 | -1 | -1 | -1 | -1 | -1 | -1 | -1 | -1 | -1 | -1 | -1 | -1 | -1 | -1 | -1 | -1 | -1 | -1 | -1 | -1 | -1 | -1 | -1 | -1 | 0 | 0 | 1 | 1 | 1 | 1 | 1 | 1 | 1 | 0.33 |
| $\Delta T_l$ | -1 | -1 | -1 | -1 | -1 | -1 | -1 | -1 | -1 | -1 | -1 | -1 | -1 | -1 | -1 | -1 | -1 | -1 | -1 | -1 | -1 | -1 | -1 | -1 | -1 | -1 | -1 | 0 | 0 | 1 | 1 | 1 | 1 | 1 | 1 | 1 | 0.20 |
| $\omega_3$ | -1 | -1 | -1 | -1 | -1 | -1 | -1 | -1 | -1 | -1 | -1 | -1 | -1 | -1 | -1 | -1 | -1 | -1 | -1 | -1 | -1 | -1 | -1 | -1 | -1 | -1 | -1 | -1 | -1 | 0 | 0 | 1 | 1 | 1 | 1 | 1 | 0.31 |
| $\theta$ | -1 | -1 | -1 | -1 | -1 | -1 | -1 | -1 | -1 | -1 | -1 | -1 | -1 | -1 | -1 | -1 | -1 | -1 | -1 | -1 | -1 | -1 | -1 | -1 | -1 | -1 | -1 | -1 | -1 | 0 | 0 | 1 | 1 | 1 | 1 | 1 | 0.23 |
| $\phi$ | -1 | -1 | -1 | -1 | -1 | -1 | -1 | -1 | -1 | -1 | -1 | -1 | -1 | -1 | -1 | -1 | -1 | -1 | -1 | -1 | -1 | -1 | -1 | -1 | -1 | -1 | -1 | -1 | -1 | -1 | -1 | 0 | 1 | 1 | 1 | 1 | 0.29 |
| $\beta_1$ | -1 | -1 | -1 | -1 | -1 | -1 | -1 | -1 | -1 | -1 | -1 | -1 | -1 | -1 | -1 | -1 | -1 | -1 | -1 | -1 | -1 | -1 | -1 | -1 | -1 | -1 | -1 | -1 | -1 | -1 | -1 | -1 | 0 | 1 | 1 | 1 | 0.00 |
| $\Delta T_g$ | -1 | -1 | -1 | -1 | -1 | -1 | -1 | -1 | -1 | -1 | -1 | -1 | -1 | -1 | -1 | -1 | -1 | -1 | -1 | -1 | -1 | -1 | -1 | -1 | -1 | -1 | -1 | -1 | -1 | -1 | -1 | -1 | -1 | 0 | 1 | 1 | 0.00 |
| $\delta$ | -1 | -1 | -1 | -1 | -1 | -1 | -1 | -1 | -1 | -1 | -1 | -1 | -1 | -1 | -1 | -1 | -1 | -1 | -1 | -1 | -1 | -1 | -1 | -1 | -1 | -1 | -1 | -1 | -1 | -1 | -1 | -1 | -1 | -1 | 0 | 1 | 0.00 |
| $T_{gr}$ | -1 | -1 | -1 | -1 | -1 | -1 | -1 | -1 | -1 | -1 | -1 | -1 | -1 | -1 | -1 | -1 | -1 | -1 | -1 | -1 | -1 | -1 | -1 | -1 | -1 | -1 | -1 | -1 | -1 | -1 | -1 | -1 | -1 | -1 | -1 | 0 | 0.00 |

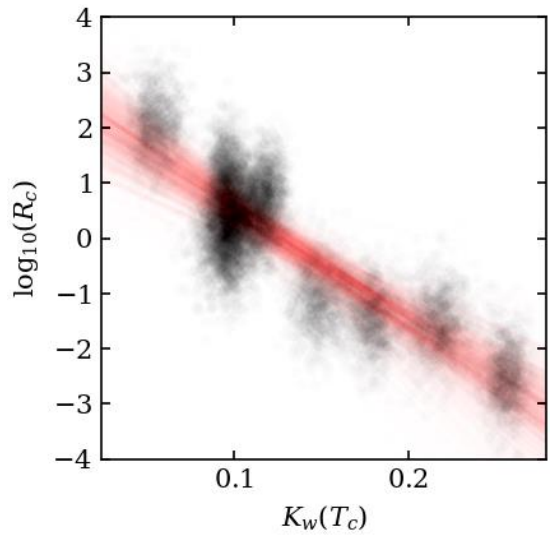
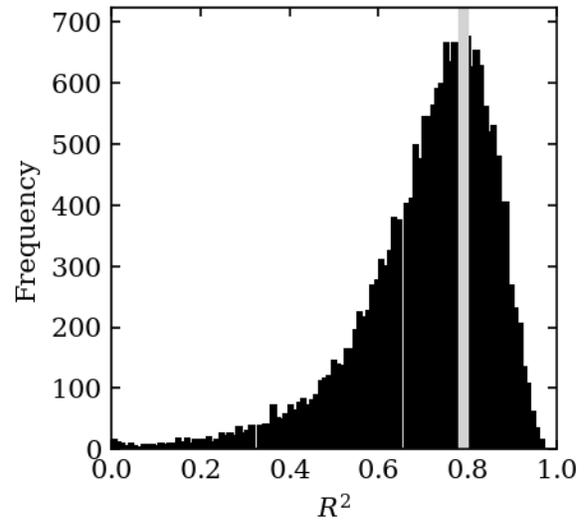
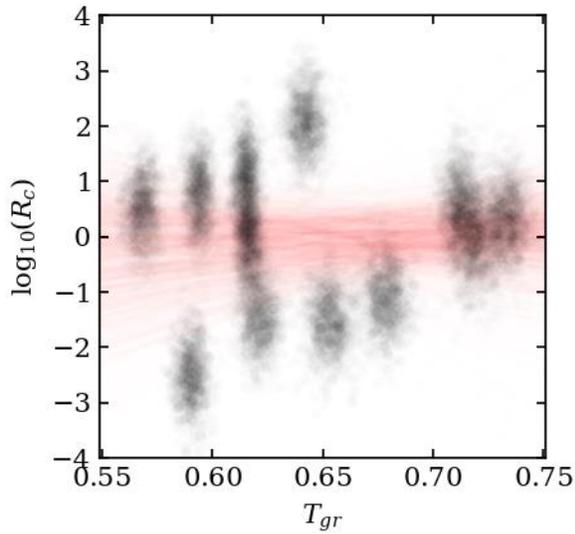
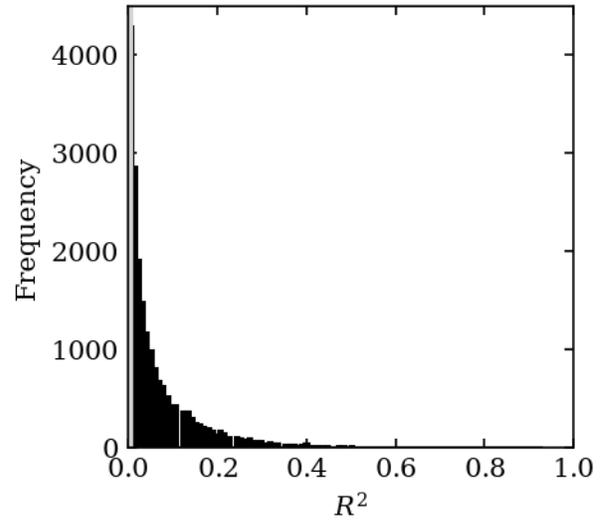

(a)                                                       (b)

(c)                                                       (d)

Figure 1 - Bootstrap distribution plot (see text for definition) and respective distribution of $R^2$ for the parameters (a) $K_w(T_c)$ and (c) $T_{gr}$, and the $R^2$ distribution for the parameters (b) $K_w(T_c)$ and (d) $T_{gr}$. The vertical gray lines in the $R^2$ plots indicate the modes of the distributions.

The bootstrap distribution plot was constructed by plotting the bootstrap samples (GS as the independent variable and GFA as the dependent variable) as black circles, which are almost transparent. The position of these circles can be revealed only if there is a reasonable overlap. The more circles overlap, the darker and more frequent that particular combination of GS and GFA is. This is a way to obtain visual information on the distribution of the bootstrap samples that were drawn.

In a similar fashion, the linear regression for each bootstrap sample is plotted as an almost transparent red line. The same reasoning used for the circles also applies for these red lines. Thus, the more saturated colors represent the places with more overlapping red lines. Observe that the red color in Figure 1a is much more saturated and contained in a much smaller area of the plot compared with Figure 1c. This is because the parameter $K_w(T_c)$ is much more robust at predicting GFA than $T_{gr}$. As complementary information, the distribution of the $R^2$ values are shown in Figure 1b and d for $K_w(T_c)$ and $T_{gr}$, respectively.

Finally, in Section S3 of the supplementary material we compare the ranking by the Wilcoxon test and the ranking by a simple correlation analysis ($R^2$) without using bootstrap sampling. Table S2 shows the R² from a simple correlation analysis, hence the reader can compare it with the more robust analysis made in this work. As a general rule, the higher the Wilcoxon rank, the higher the correlation rank is. However, these ranks are *not* exactly the same, which shows that good correlation does not necessarily imply equally good (statistical) power of prediction. The R² distribution plots for the other parameters, similarly to the plots shown in Figure 1b and Figure 1d, are shown in Section S5 of the supplementary material.

## 4.1 The best GS parameters to predict $R_c$

Table 4 shows that only parameters $K_w$, $K_H$, $\gamma$, H', $\Delta T_{rg}$, and $K_{cr}$ have the mode of $R^2$ greater than or equal to 0.70. The next on the list has a mode less than 0.65. The simple linear correlation coefficients for these parameters span from 0.75 to 0.86. While this is in an arbitrary threshold, here we deemed these parameters as being the best statistical predictors of $R_c$. In the following paragraphs, we will discuss the main features of these 6 parameters.

The $K_w = (T_c-T_g)/T_l$ parameter, proposed by Weinberg, includes a normalization by $T_l$ [23] of the $\Delta T_g (T_c) = T_c-T_g$ parameter, which is taken by several authors as a measure of glass stability, e.g. [17,31]. In Weinberg´s work, however, $T_c$ was calculated using crystallization models, instead of being a datum from a DTA or DSC experiment.

The $K_H = (T_x-T_g)/(T_m-T_x)$ parameter proposed by Hrübý [13] was deduced based on the consideration that the GFA is directly proportional to the glass stability. He considered a DTA analysis as a reproduction, in a slower rate, of the quenching step in the glass making process. He observed that $T_x$ is located at higher temperatures for glasses that are more easily obtained, eventually as high as the melting temperature. In this case, crystallization would be difficult and a glass would be very easy to make. The $T_x-T_g$ term was thus related to these considerations. Hrübý also affirmed that a short $T_m-T_x$ interval indicates that the crystalline phase formed at $T_x$ has a relatively low melting point, which is considered a positive feature for easy vitrification. Therefore, $T_m-T_x$ was considered to be inversely proportional to the GFA.

The H' = $(T_x-T_g)/T_g$ parameter is just another normalized form of the $\Delta T_g = (T_x-T_g)$. It is not clear who the authors were of these two parameters, but they were used by Saad and Poulan in 1987 [31]. Saad and Poulain suggested that the difference between $T_x$ (or $T_c$) and $T_g$ is related to the GFA, because they observed a maximum value for the best glass forming composition in their study. Normalization by $T_g$ would enable the comparison between glasses of different systems. This parameter only depends on two characteristic temperatures.

The $\gamma = T_x/(T_g+T_l)$ parameter, proposed by Lu and Liu [34], has two different derivations. The most simple of them, was published in [94], and considers the TTT curve for homogeneous nucleation. Interestingly, the basis of the current work is heterogeneous nucleation, which is quite different from the considerations of Lu and Liu. The authors suggested that two aspects of the liquid stability should be considered: the liquid stability at equilibrium state and the liquid stability at the (metastable) supercooled state. When two glasses have the same $T_g$, the one with the lowest $T_l$, has the highest liquid phase stability. For glasses having the same $T_l$ but different $T_g$, the one with the lowest $T_g$ has the highest liquid stability at the metastable state. To compare glass-forming liquids with distinct $T_l$ and $T_g$, one can use ½ $(T_g+T_l)$. The smaller $(T_g+T_l)$, the higher the stability of the supercooled liquid. Now, considering that the TTT curves for all glasses have the same shape, $T_x$ would be the temperature where a heating thermal experiment crosses the lower part of the TTT nose. Considering the same shape of the TTT curve for all glasses, $T_x$ is proportional to the time in the nose of the TTT curve. For glasses with the same $(T_g+T_l)$, a higher $T_x$ means a lower $R_c$ and, therefore, the higher the $\gamma$ parameter, the higher the GFA.

The $\Delta T_{rg} = [(T_x-T_g)/(T_l-T_g)]$ and $K_{cr} = [(T_l-T_x)/(T_l-T_g)]$ parameters are equivalent ($K_{cr} = 1 - \Delta T_{rg}$). The origin of this parameter is not clear, but it appears in the work of Du and Huang (2008) [17], listed as one of the parameters that has already been used in previous GFA studies. The $K_{cr}$ was proposed by Polyakova (2015) [55] who deduced it by starting with the Hrübý parameter. Polyakova observed that the $K_H$ is a non-linear function of $T_x$, which is responsible for its low sensitivity when $T_x$ is near $T_g$, and highly sensitive when $T_x$ is near $T_l$. By substituting the $T_l-T_x$ by $T_l-T_g$ in the denominator, the $K_{cr}$ becomes a linear function of $T_x$ and can assume values between 0 - for glasses with outstanding GFA - to 1 - for glasses with really poor GFA.

All the best GS parameters require 3 characteristic temperatures of the material, H' being a notable exception. Interestingly, when not tied, the parameters $K_w$, $K_H$, and $\gamma$ performed better with $T_c$ than their counterparts with $T_x$. This result makes sense given that the position of the crystallization peak is where the rate of crystallization is maximum, a key temperature to be avoided to vitrify any substance. $T_c$ is also better defined in DTA/DSC traces than $T_x$. It is important to note that to determine the characteristic temperatures; we analyzed samples having similar particle size distributions and the same heating rate. This is relevant for comparison studies because these two process variables significantly affect $T_x$ and $T_c$, as observed by Nascimento et al. [24], who observed the influence of such parameters considering fine and coarse grains.

# 5 Summary and Conclusions

We used bootstrap sampling and the Wilcoxon test to find out which glass stability parameter can statistically predict the glass forming ability. To use this methodology, we collected or measured the necessary data for twelve stoichiometric oxide glasses that underwent predominant heterogeneous nucleation. To the best of our knowledge, such an extensive statistical test was used for the first time to check the capacity of 35 GS parameters to access the GFA.

We found that some GS parameters predict the GFA of oxide glasses quite well whereas most perform poorly. Parameter **$K_w$** was the best, closely followed by the $K_H$, $\gamma$, $H'$, $\Delta T_{rg}$, and $K_{cr}$. Our results corroborate previous reports carried out using a smaller number of glasses, much less GS parameters, and less rigorous statistics. We also found that using $T_c$ instead of $T_x$ improved the predictive power of these parameters. Finally, the only parameter considered here that predicts the GFA without requiring the production of a glass piece to be used, *Jezica*, ranked reasonably well in our analysis, even though it does not rely on any crystallization information to predict GFA.

## Acknowledgements

This study was financed by the São Paulo State Research Foundation (FAPESP), grant numbers 2017/12491-0 (DRC) and 2013/07793-6 (CeRTEV), by the National Council for Scientific and Technological Development (CNPq), grant number 141107/2016-2, and by the Coordenação de Aperfeiçoamento de Pessoal de Nível Superior - Brasil (CAPES) - Finance Code 001. We are grateful to Marcio L. Nascimento and Eduardo B. Ferreira for their valuable suggestions.

# Supplementary material

## S1. DTA analysis

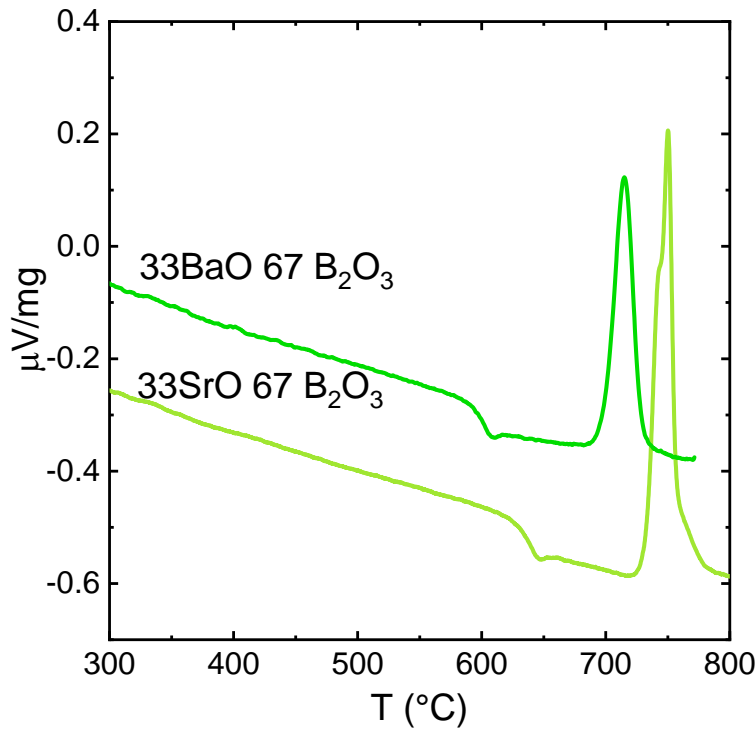

Figure S 1 - Thermal analysis of the borate glasses produced in this work.

## S2. Statistical method explained in detail

Here we discuss *bootstrapping*, the widespread statistical method [90–92] used to answer the main question of this work: *"which GS parameters best predict the GFA?"* We divided this method into five steps: resampling, adding noise, calculation, correlation, and prediction.

*Resampling* is the process of creating a "bootstrap sample" by randomly sampling the original dataset with reposition. One constraint is that the bootstrap sample must have the same size of the original dataset. In our case, the original dataset contains twelve glasses, as shown in Table 2. Thus, any bootstrap sample that is generated must also contain twelve glasses. Due to the random nature of the sampling process with reposition, the bootstrap sample will most probably be a different set of glasses than the original dataset, with some glasses occurring more than once and some glasses not appearing at all. We will call "unselected glasses" the glasses that were not selected during sampling.

*Adding noise* is the process of simulating the uncertainty of experimental measurements. Each entry in a bootstrap sample and the "unselected glasses" will incorporate some random amount of noise after the resampling step. The noise may be *absolute*, for which a value drawn from a normal distribution with zero mean and standard deviation specified in Table Table S1 was added to the parameter; or *relative*, for which a value drawn from a normal distribution with a mean of 1 and specified standard deviation is Table multiplied by the parameter. We only interfered in this random process for the cases where, by chance, the temperature of maximum crystal growth ($T_{Umax}$) was greater than or equal to the melting temperature, which is physically absurd. In these cases, we recomputed all the noises within the bootstrap sample until no such forbidden situation occurred.

Table S1 - Type, mean, and standard deviation of the noise that was added to the parameters drawn during the bootstrap sampling. Absolute noises were added to the parameter and relative noises multiplied by it (see text).

| Parameter | Type of noise | Mean | Standard deviation |
|---|---|---|---|
| $T_g$ | Absolute | 0 | 5 |
| $T_l$ | Absolute | 0 | 5 |
| $T_c$ | Absolute | 0 | 8 |
| $T_x$ | Absolute | 0 | 8 |
| $T_{Umax}$ | Absolute | 0 | 8 |
| $U_{max}$ | Relative | 1 | 0.02 |
| $\log_{10}(N_s)$ | Absolute | 0 | 1 |
| $\log_{10}(\eta)$ | Absolute | 0 | 0.2 |

*Calculation* is the process of computing $R_c$ and the value of all the GS parameters for the glasses in a bootstrap sample and the respective "unselected glasses." Computation of $R_c$ was done using Eq. 2, with $X_s = 10^{-2}$ and $N_s = 10^3$ (we also considered a noise in the value of $\log_{10}(N_s)$, as shown in Table ). Viscosity at the liquidus temperature was only needed for the Jezica parameter, and it was computed using Eq. 1. The equations for all the GS parameters are shown in Table 1.

*Correlation* is the process of performing linear regressions using the bootstrap sample data. One linear regression for each GS parameter was obtained considering the GS parameter as the independent variable and $\log_{10}(R_c)$ as the dependent variable. In our case, because we are analyzing 35 GS equations, we obtained 35 linear equations for each bootstrap sample, each one of these regressions using a different GS parameter as the independent variable. We also computed the coefficient of determination for each linear regression that was performed.

*Prediction* is the process of predicting the value of $\log_{10}(R_c)$ of the "unselected glasses" by using the GS parameters of the "unselected glasses" and the linear equations obtained in the previous step. Having the predicted $\log_{10}(R_c)$ for each GS parameter and the calculated $\log_{10}(R_c)$ of these "unselected glasses", we computed the absolute residual of the prediction for each GS parameter, which is the absolute difference between the predicted and calculated $\log_{10}(R_c)$.

The five steps of the method, discussed in the previous paragraphs, were done 10,000 times, each time storing in individual lists the values of the coefficient of determination obtained in the *Correlation* step and storing the absolute residual of prediction obtained in the *Prediction* step. Each GS parameter had its own list of coefficient of determination and residuals.

With the list of the coefficient of determination, we computed the mode of this coefficient for each GS parameter. The mode of $R^2$ is reported in Table 4.

With the list of residuals, we can test which GS parameter yields the lowest residuals, thus having the best prediction power. To do this we used the non-parametric Wilcoxon test with a statistical confidence of 99%. This test is done in pairs of GS parameters. First, we tested the null hypothesis if the difference between the absolute residuals of a pair of GS parameters followed a symmetric distribution around zero. If we cannot reject the null hypothesis, then we cannot claim that there is a statistical difference in the prediction of $\log_{10}(R_c)$ for the considered pair of parameters. However, when the null hypothesis was rejected, we ran the Wilcoxon test again having as a null hypothesis the absolute residuals of one of the GS parameters are lower than the other. By performing the Wilcoxon test with each possible pair of GS parameters, we were able to check which (if any) GS parameters are best to predict the $\log_{10}(R_c)$ as the parameters that yield the lowest absolute residuals.

## S3. Simple correlation analysis *versus* Wilcoxon test

Table S2 – Correlation coefficients, $R^2$, from a linear correlation analysis between the GS parameters and $R_c$.

| Parameter | $K_w(T_c)$ | $K_w(T_x)$ | $\gamma(T_c)$ | $H'(T_x)$ | $H'(T_c)$ | $K_H(T_c)$ | $\Delta T_{rg}$ | $K_{cr}$ | $K_H(T_x)$ | $\gamma_c$ | $K_M$ | $\omega_2$ |
|---|---|---|---|---|---|---|---|---|---|---|---|---|
| $R^2$ | 0.86 | 0.83 | 0.80 | 0.80 | 0.80 | 0.79 | 0.77 | 0.77 | 0.77 | 0.75 | 0.69 | 0.66 |

| Parameter | $\beta$ | $\gamma_m$ | $\omega_0$ | $\gamma(T_x)$ | Jezica | $K_{w2}(T_x)$ | $\beta_2$ | $\Delta T_c$ | $G_p$ | $\Delta T_x$ | $K_S(T_x)$ | $\omega_1$ |
|---|---|---|---|---|---|---|---|---|---|---|---|---|
| $R^2$ | 0.64 | 0.64 | 0.62 | 0.61 | 0.60 | 0.60 | 0.60 | 0.60 | 0.58 | 0.58 | 0.57 | 0.54 |

| Parameter | $\alpha(T_c)$ | $K_S(T_c)$ | $\xi$ | $\alpha(T_x)$ | $\omega_3$ | $\phi$ | $\Delta T_l$ | $\Delta T_g$ | $\delta$ | $T_{rg}$ | $\theta$ | $\beta_1$ |
|---|---|---|---|---|---|---|---|---|---|---|---|---|
| $R^2$ | 0.52 | 0.51 | 0.46 | 0.32 | 0.30 | 0.28 | 0.25 | 0.03 | 0.01 | 0.01 | 0.00 | 0.00 |

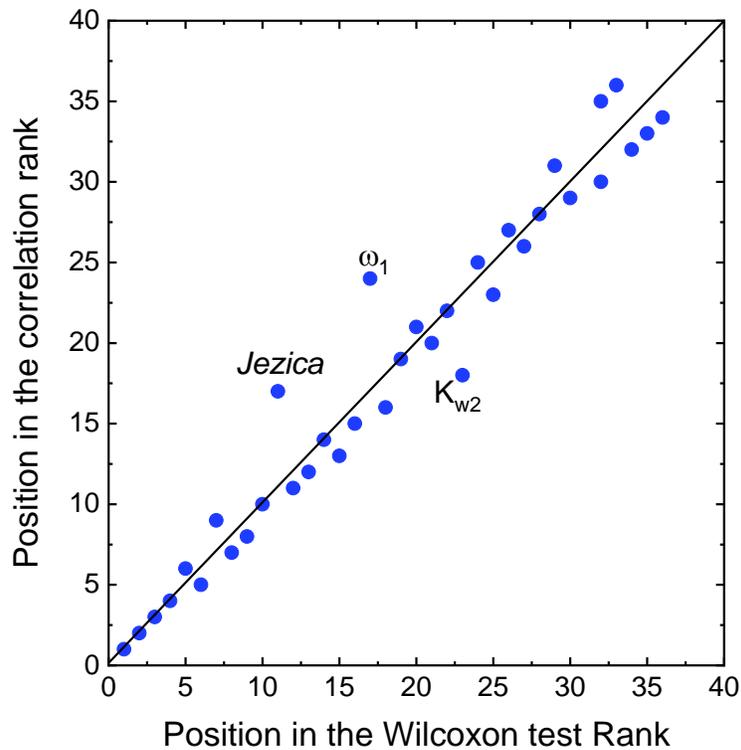

Figure S 2 - Position in the simple correlation ($R^2$) rank vs. position in the Wilcoxon test rank.

The data points for which the diagonal line passes through have the same position in both ranks. In general, they are well correlated; the Jezica, $\omega_1$, and $K_{w2}$ show the greatest difference between the two types of tests.

## S4. Python code

The supplementary Python code used in this work is licensed under GPL-3.0 and is freely available in the GitHub repository https://github.com/drcassar/supp_code_glass_stability.

# S5. R² plots

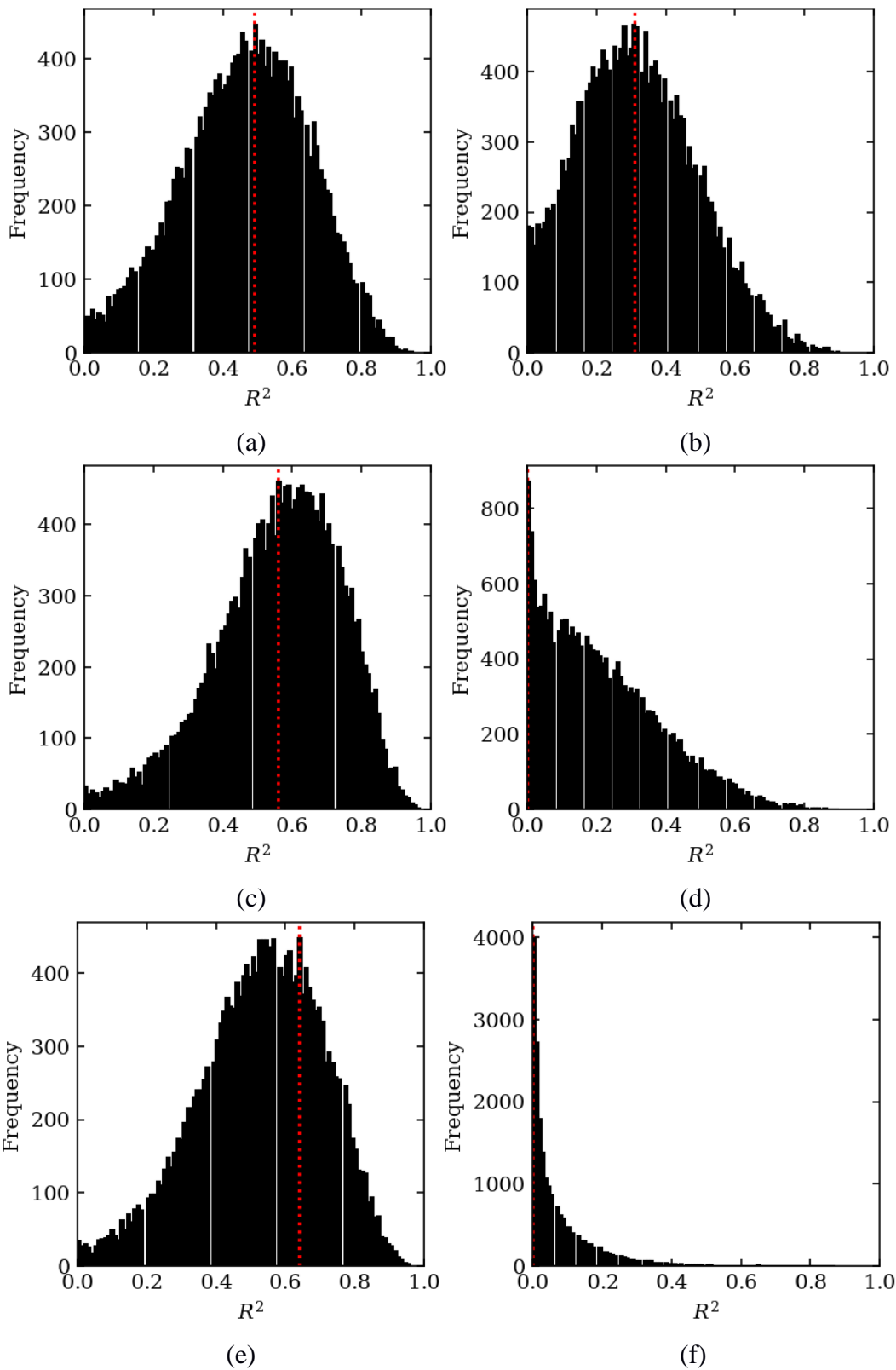

Figure S3 - Distribution of $R^2$ for the parameters (a) $\alpha(T_c)$, (b) $\alpha(T_x)$, (c) $\beta$, (d) $\beta_1$, (e) $\beta_2$ and (f) $\delta$.

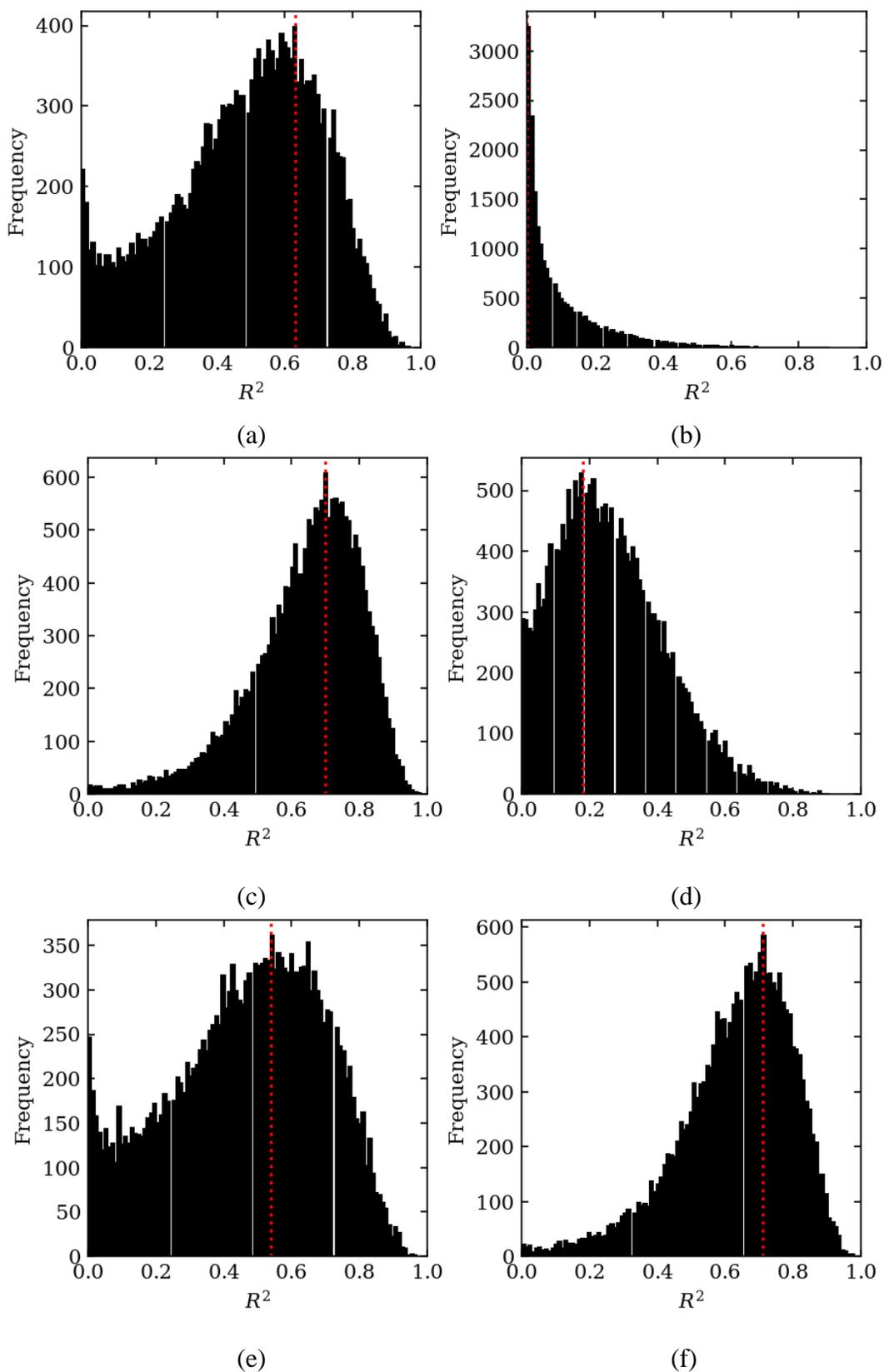

Figure S4 - Distribution of $R^2$ for the parameters (a) $\Delta T_c$, (b) $\Delta T_g$, (c) $\Delta T_{gr}$, (d) $\Delta T_l$, (e) $\Delta T_x$ and (f) $\gamma_c$.

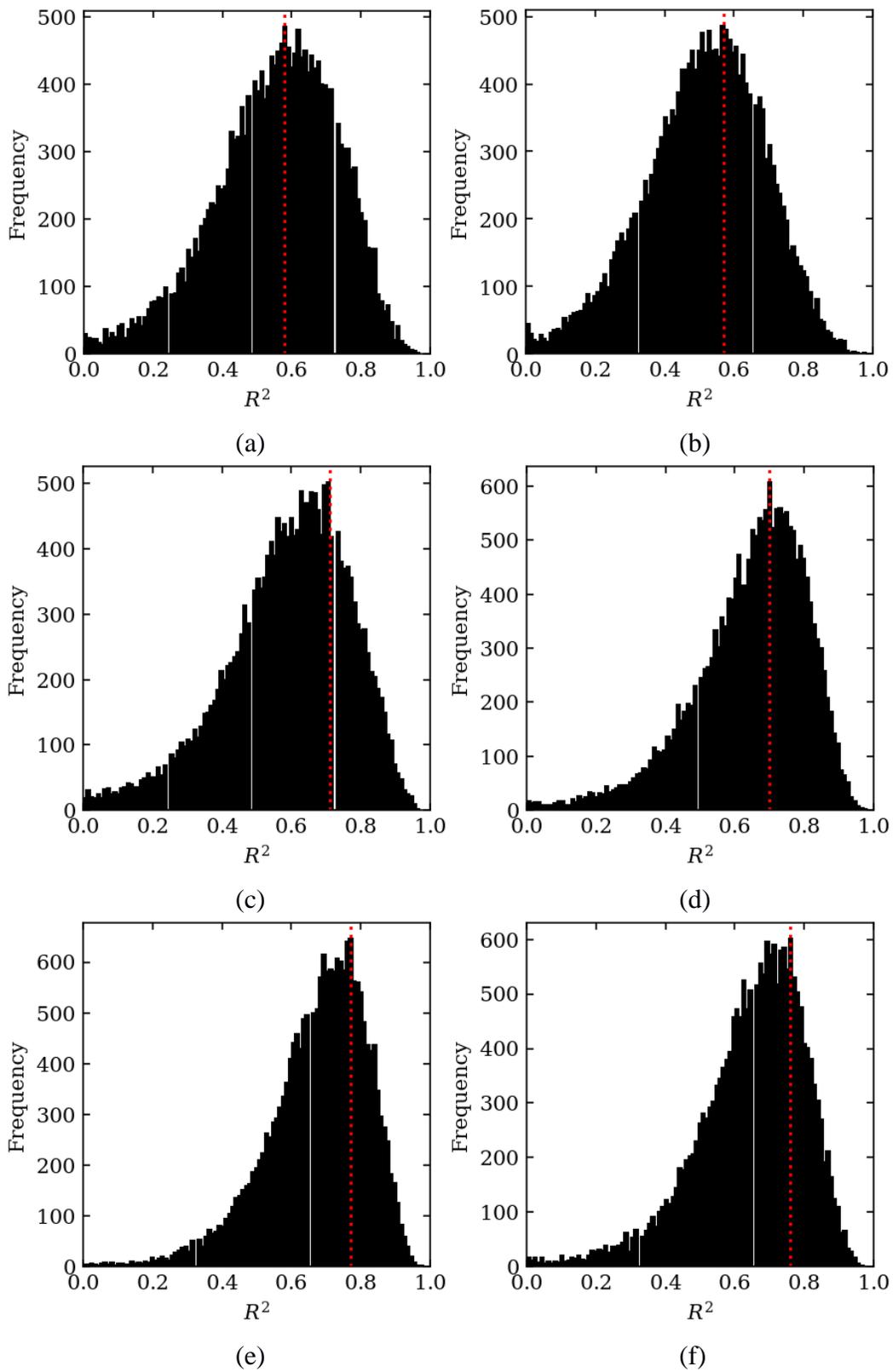

Figure S5 - Distribution of $R^2$ for the parameters (a) $\gamma_m$, (b) $G_p$, (c) *Jezica*, (d) $K_{cr}$, (e) $K_H(T_c)$ and (f) $K_H(T_x)$.

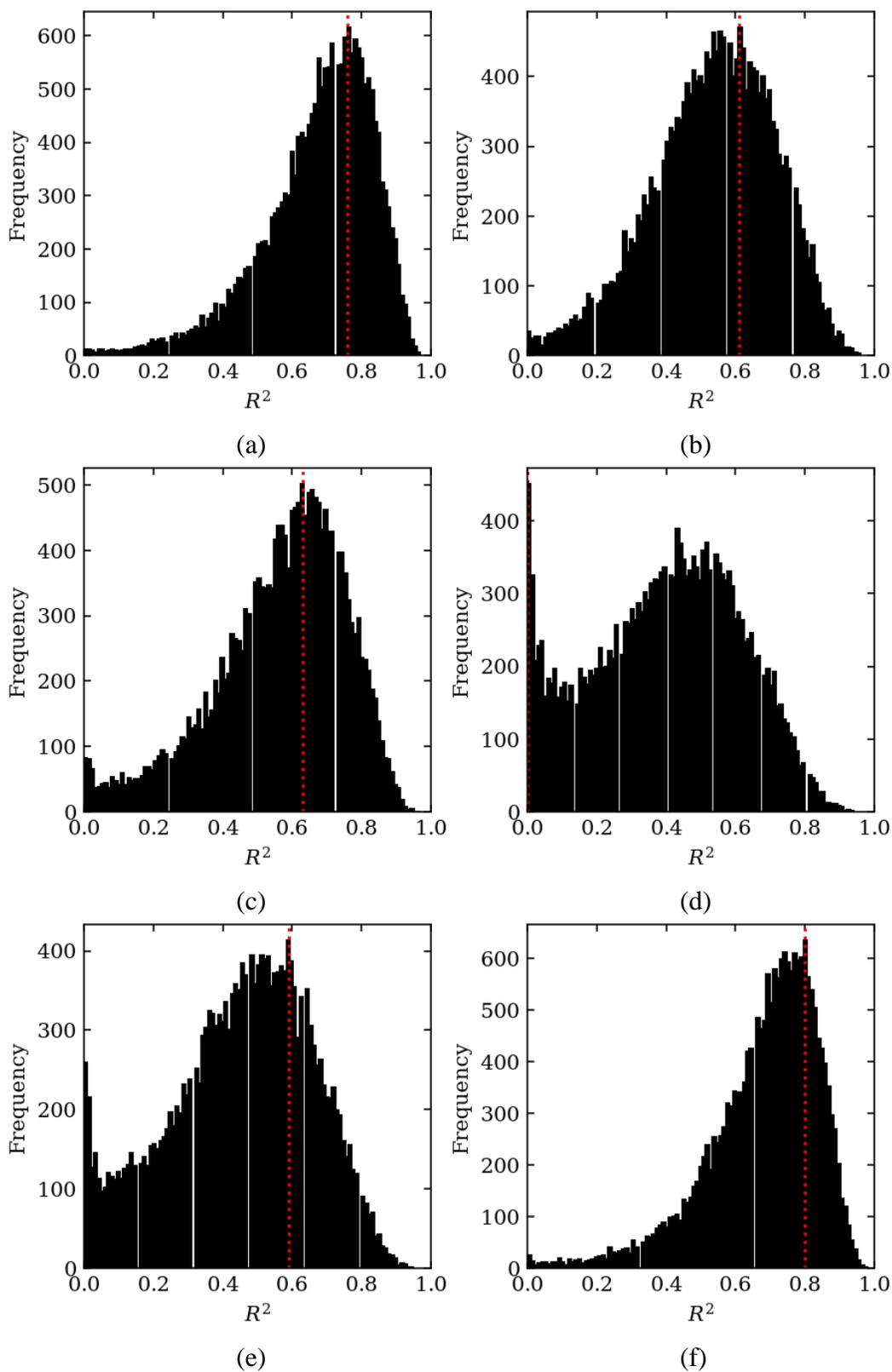

Figure S6 - Distribution of $R^2$ for the parameters (a) $\gamma$ ($T_c$), (b) $\gamma$ ($T_x$), (c) $K_m$, (d) $K_S(T_c)$, (e) $K_S(T_c)$ and (f) $K_w(T_x)$.

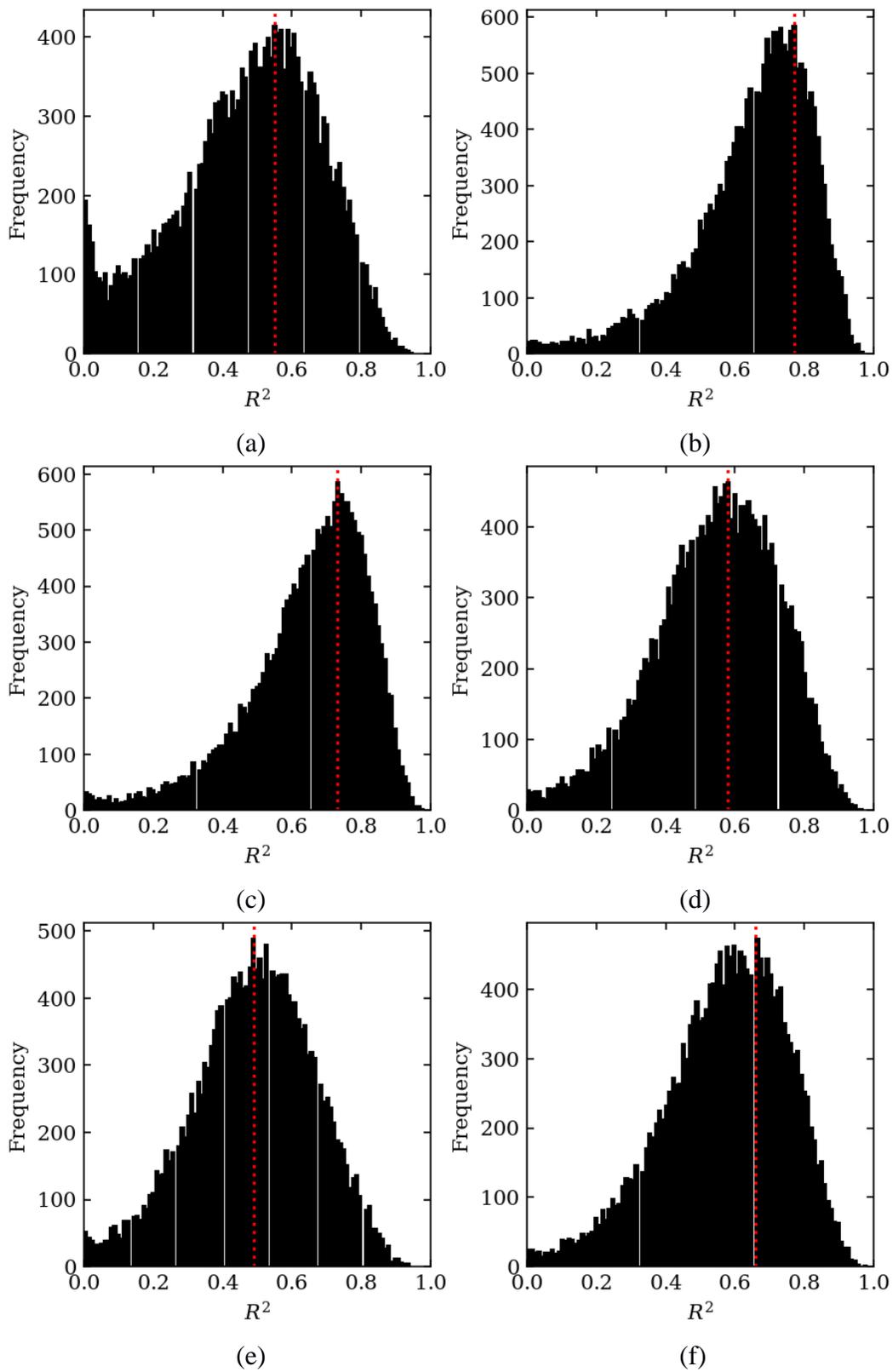

Figure S7 - Distribution of $R^2$ for the parameters (a) $K_{w2}$, (b) $H'(T_c)$, (c) $H'(T_x)$, (d) $\omega_0$ (e) $\omega_1$ and (f) $\omega_2$.

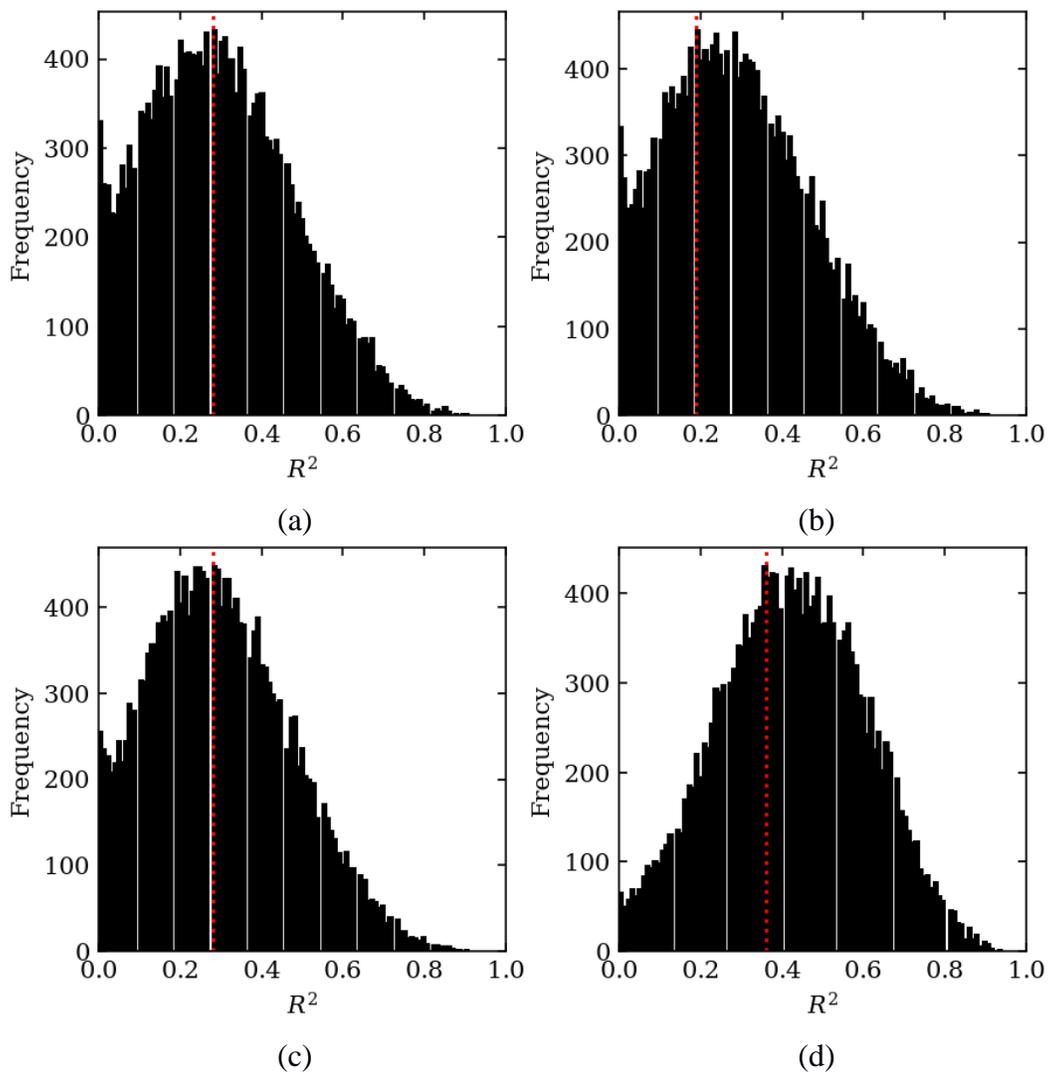

Figure S8 - Distribution of $R^2$ for the parameters (a) $\omega_3$, (b) $\phi$, (c) $\theta$ and (d) $\xi$.